\documentclass[preprint,showpacs,aps]{revtex4}
\usepackage{graphicx}
\usepackage{bm}
\usepackage{psfig}

\newcommand{\be}{\begin{equation}}
\newcommand{\ee}{\end{equation}}
\newcommand{\bea}{\begin{eqnarray}}
\newcommand{\eea}{\end{eqnarray}}
\newcommand{\ba}{\begin{array}}
\newcommand{\ea}{\end{array}}
\newcommand{\bi}{\begin{itemize}}
\newcommand{\ei}{\end{itemize}}
\newcommand{\ben}{\begin{enumerate}}
\newcommand{\een}{\end{enumerate}}

\begin{document}
\title{Large-Eddy Simulations of Fluid and Magnetohydrodynamic Turbulence Using Renormalized Parameters} 
\author{Mahendra\ K.\ Verma and Shishir Kumar} 
\affiliation{Department of Physics, Indian Institute of Technology, Kanpur -- 208016, INDIA}
\date{March 7, 2003}

\begin{abstract}
  In this paper a procedure for large-eddy simulation (LES) has been
  devised for fluid and magnetohydrodynamic turbulence in Fourier
  space using the renormalized parameters.  The parameters calculated
  using field theory have been taken from recent papers by Verma
  \cite{MKV:MHD_PRE,MKV:MHDRG}.  We have carried out LES on $64^3$
  grid.  These results match quite well with direct numerical
  simulations of $128^3$.  We show that proper choice of parameter is
  necessary in LES.
\end{abstract}

\vspace{1cm}

\maketitle

Turbulence is one of the most difficult and unsolved problems of
classical physics.  To probe the complex dynamics of turbulence, one
often resorts to computer experiments, known as Direct Numerical
Simulations (DNS).  Since multiple scales are involved in turbulence,
DNS of turbulence is a very expensive task in terms of both computer
time and memory, even in modern computers. For example, a
pseudo-spectral simulation by Gotoh~\cite{Goto:DNS} on $1024^3$ grid
using vector parallel Fujitsu VPP5000/56 with 32 processors took 500
hours of computer time, and required 8 Gigabytes of memory per
processor.  To reduce the required computer time and memory space, an
ingenious technique called large-eddy simulation (LES) has been
developed (see review article by Metais~\cite{Meta:rev} and references
therein).

Basic idea of LES is to resolve only the large scales of turbulent
flow.  The effect of smaller scale interactions are modeled
appropriately using the existing theories.  In turbulence, Fourier
modes of different scales interact with each other.  Kolmogorov
provided an important model of turbulence in which the interactions
effectively yield a constant energy flux from large scales to
intermediate scales, and then to small scales.  When we observe
Fourier modes up to certain length scale $l$ in the intermediate
range, the modes with scales less than $l$ act as a sink of energy.
According to Kolmogorov's theory, the amount of sink should be equal
to the energy flux. In LES, the large scales up to $l$ are resolved by
using eddy viscosity at cutoff scale $l$, where energy is drained.
Analysis of turbulence using renormalization groups (RG) shows that
the above modeling is possible. LES uses this idea to analyze
large-scale dynamics of turbulence.

Renormalization Group (RG) is a popular tool used by physicists to
solve problems with multiple scales.  Since turbulence involves
multiple scales, RG has been applied successfully to turbulence
\cite{FNS,YakhOrsz,McCo:book}. In Wilson's Fourier space RG scheme,
Fourier space is divided into many shells.  The nonlinear interactions
among various shells are computed using first-order perturbation
theory, that yields an effective viscosity, called renormalized or
eddy viscosity, at any scale.  The renormalized viscosity is found to
be wavenumber ($k$) dependent.  McComb and Watt~\cite{McCoWatt}
computed the renormalized viscosity using `self-consistent' RG
procedure. When the cutoff wavenumber $k_C$ is in the inertial range,
the renormalized viscosity is given by
\begin{equation}
\nu_r(k_C)  =  (K)^{1/2} \Pi^{1/3} k_C^{-4/3} \nu^*    
         \label{eqn:nu}
\end{equation}
where $\Pi$ is the energy flux, $K$ is Kolmogorov's constant, and
$\nu^*$ a parameter.  McComb and Watt~\cite{McCoWatt} found $\nu_*
\approx 0.50$ and  $K \approx 1.62$. Verma
\cite{MKV:MHD_PRE} also computed the above quantities using a refined
technique and found $\nu_* \approx.38$ and $K \approx 1.6$.

Zhou and Vahala~\cite{ZhouVaha88,ZhouVaha93} developed an alternative
recursive-renormalization-group theory for turbulence modeling. In
their calculation they find backscatter of energy from small scales to
large scales, and a cusp in renormalized viscosity near $k_C$.  
These features are attributed to triple correlations, which has not been
accounted for in McComb and Watt's calculations.  Recently Schilling and
Zhou~\cite{SchiZhou} have addressed the above problem using
eddy-damped quasinormal Markovian (EDQNM) closure model.  In the current
paper we neglect backscatter.

McComb~\cite{McCo:book} had proposed that the renormalized viscosity
$\nu(k_C)$ could be used as effective viscosity for LES, however, this
calculation had not been done till date. Earlier, the spectral eddy
viscosity $\nu_t(k|k_C)$ has been used for LES in EDQNM formalism
(see~\cite{Lesi:book}).  In this scheme,

\begin{equation}
\nu_t(k|K_C) = 0.441 K^{-3/2} \left[ \frac{E(k_C)}{k_C} \right]^{1/2}
                f(k/k_C)
\label{eqn:nu_t}
\end{equation}
where $f(x)$ is a nondimensional function which tends to 1 as $x$
approaches 0.  Comparing Eqs.~(\ref{eqn:nu},\ref{eqn:nu_t}), we find
that their dependence on Kolmogorov's constant is different.  In
Eq.~(\ref{eqn:nu_t}), if we assume that $E(k_C)$ follows Kolmogorov's
spectrum and $K=1.6$, then the constant multiplying $\Pi^{1/3}
k_C^{-4/3}$ is 0.27.  In contrast, in Eq.~(\ref{eqn:nu}) the same
quantity is $\sqrt{K} \nu^* \approx 0.48$.  As it will be shown
in the later part of the paper, the choice of constant is quite
crucial in LES.  We find that $\nu_r(k_C)$ of Eq.~(\ref{eqn:nu})
yields better numerical results compared to $\nu_t(k|k_C)$ of
Eq.~(\ref{eqn:nu_t}).  We believe that the calculation of renormalized
viscosity is theoretically more sound than the calculation of spectral
eddy viscosity using EDQNM approximation, therefore, former is
more appropriate for LES than the later.

In this paper we perform LES of fluid turbulence using renormalized
viscosity.  We have been able to apply the same procedure to
magnetohydrodynamic (MHD) turbulence also, except that we need two
renormalized parameters: renormalized viscosity and renormalized
resistivity.  The required parameters for MHD have been recently
calculated by Verma \cite{MKV:MHD_PRE,MKV:MHDRG,MKV:MHDflux}.  The LES
calculations have been performed on $64^3$ grid, and they have been
compared with DNS results of $64^3$ and $128^3$.  As described below,
the inertial range in LES is found to be either equal or larger than
that in DNS, hence our LES model is working very well.

We solve Navier-Stokes equation in Fourier space~\cite{Orsz}:
\begin{equation}
\frac{\partial u_i ({\bf k})}{\partial t}    =  
-\nu_r(k_C) k^2 u_i ({\bf k}) -   FT\left(
        u_j \frac{\partial u_i}{\partial x_j} \right)- i k_i p({\bf k})
\label{eqn:NSk} 
\end{equation}
where $FT$ stands for Fourier transform.  We take $\nu^*$ to be equal to 0.38.

We adopt pseudo spectral method on grid size $64^3$ with $dt=10^{-4}$.
We apply Adam-Bashforth scheme to integrate the nonlinear terms, and
Crank-Nicholson's scheme for the viscous term.  We apply 2/3 rule to
eliminate the aliasing errors \cite{Orsz}.  We use Fast Fourier
Transform developed by Frigo and Johnson~\cite{fftw} for our
calculations.  Our initial condition is taken to be unit energy spread
out in wavenumber shells from 2 to 13 with an exponentially decreasing
distribution. The modes in a shell have equal energy but random
phases, and satisfy divergenceless condition. The most important
ingredient in our simulation is renormalized viscosity, which is
computed using Eq.~(1) with $k_C=32$.  Since $\Pi$ changes with time,
it is computed every 0.01 dimensionless time unit.  We use dissipation
rate for $\Pi$.  We carry out our simulation up to 50 time units.  Our
LES simulation  takes approximately 60 hours on Athlon 1.7 GHz
processor.

\begin{figure}[h]
\centerline{\mbox{\psfig{file=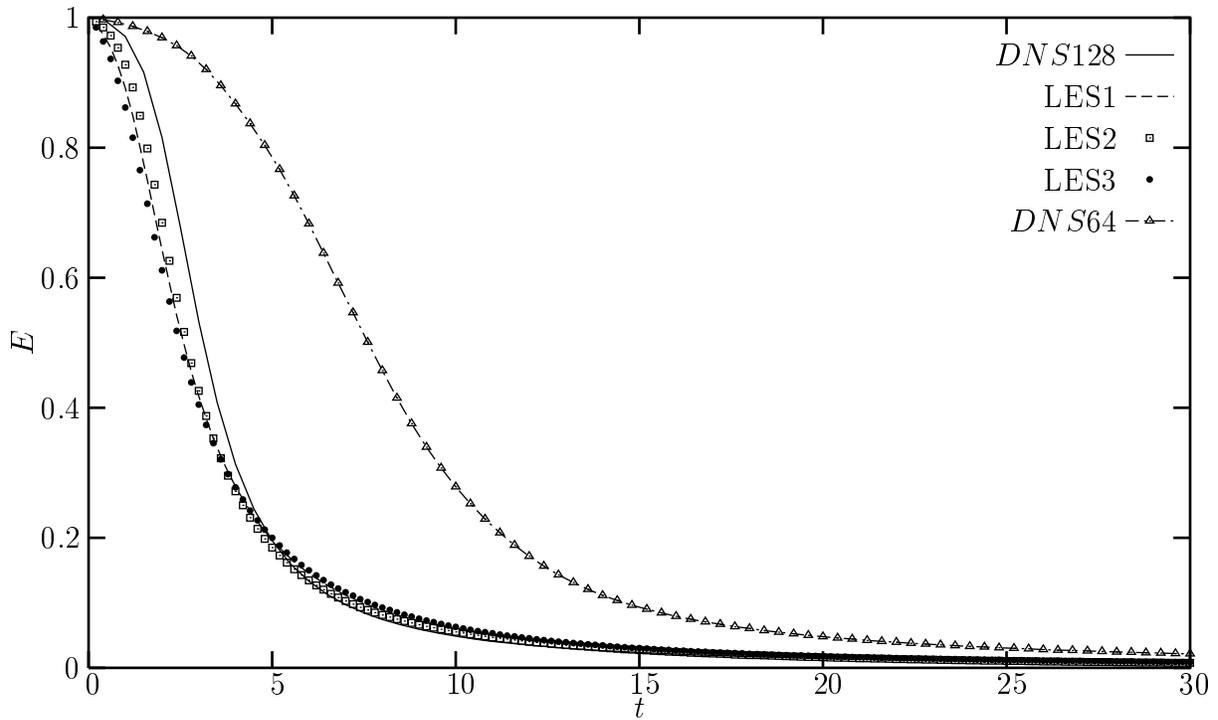,width=1.0\textwidth}}}
\caption{Temporal evolution of energy in fluid turbulence using DNS
and LES.  The figure contains Energy($E$) vs time plots for DNS128
(solid line), DNS (DNS64), and three LES runs using $\nu^*$ equal to
0.38 (LES1), 0,25 (LES2) and 0.48 (LES3).  The evolution in LES for
all the three $\nu^*$ is quite close to DNS128, but not to
DNS64.}
\label{fig:et_fluid}
\end{figure}

In Fig.~\ref{fig:et_fluid} we show the energy evolution as a function of
time for $\nu^*= 0.25,0.38,0.48$.  The $E$ vs.~$t$ plot for all three
$\nu^*$ are overlapping.  The LES results are also compared with the
standard pseudo-spectral DNS results performed on $64^3$ (DNS64) and
$128^3$ (DNS128) with identical initial condition and $\nu_0=2 \times
10^{-4} $.  In DNS we apply additional hyperviscous term $1/k_{eq}^2
k^4 {\bf u(k)}$ with $k_{eq}=9$ to overcome aliasing errors. Clearly
the energy evolution for LES matches quite well with DNS128, but
differ significantly with DNS64. Hence, our LES on $64^3$ is able to
mimic DNS of $128^3$.

\begin{figure}[h]
\centerline{\mbox{\psfig{file=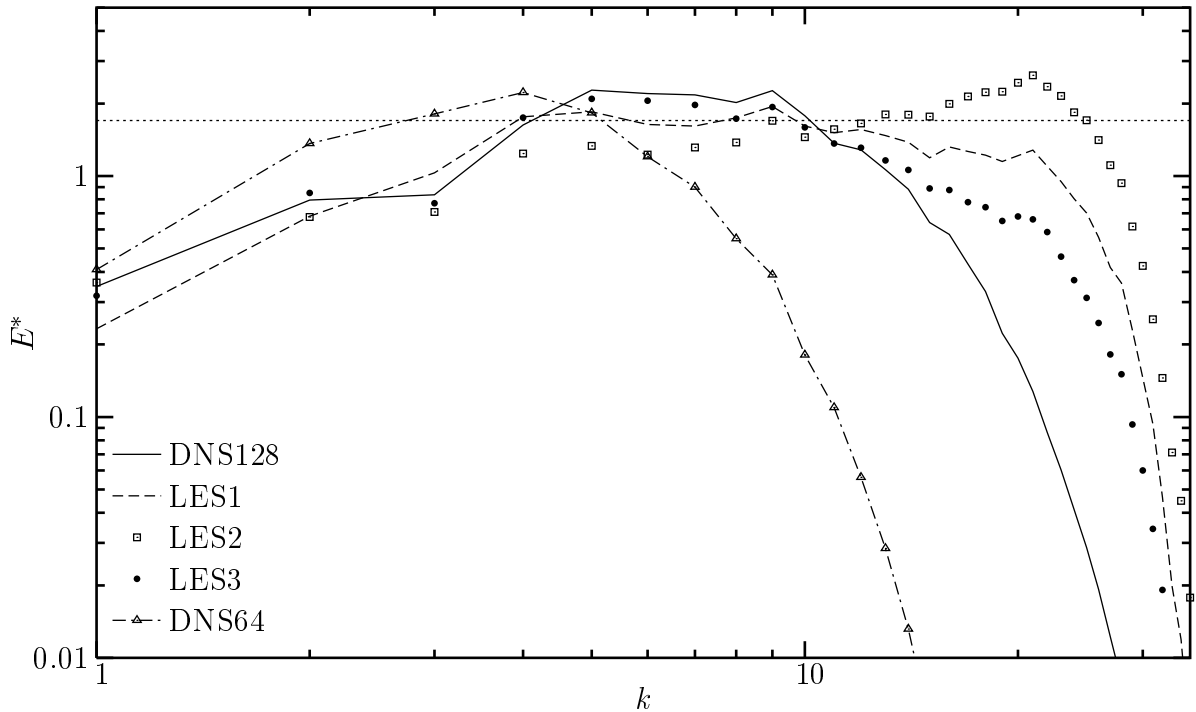,width=1.0\textwidth}}}
\caption{Energy spectrum for fluid turbulence is calculated using DNS
and LES.  The figure contains plots of normalized energy spectrum
$E'(k)=E(k)k^{5/3}\epsilon^{-2/3}$ with wavenumber $k$ after 50 time
units for DNS128, DNS64, and three LES runs using $\nu^*$ equal to 0.38 (LES1),
0,25 (LES2) and 0.48 (LES3).  We get the best inertial range for
$\nu^*=0.38$.  The Kolmogorov's constant is found to be $1.7 \pm
0.1$. DNS64 run has hardly any inertial range.}
\label{fig:ek_fluid}
\end{figure}

In Fig.~\ref{fig:ek_fluid}, we plot $E(k) k^{5/3} \Pi^{-2/3}$ vs. $k$
for DNS as well as LES.  Again the normalized spectrum of LES matches
quite well with DNS128 at small and intermediate wavenumbers. Note
that $64^3$ LES has much larger inertial range compared to $64^3$ DNS,
where it is almost absent.  We find that the wavenumber range of
inertial wavenumbers (constant with $k$) is maximum for LES with
$\nu^*=0.38$; in fact wavenumber range for LES is larger than that for
DNS128.  The energy spectrum for $\nu^*=0.25$ has a hump for large
wavenumbers (underdamped case), implying that actual $\nu^*$ value is
higher than 0.25.  The spectrum for $\nu^*=0.48$ shows overdamped
character \cite{Math:book}.  We have done DNS128 for some more
parameters.  The trend appears to show that $\nu^* \approx 0.38$ is
the most appropriate choice for LES.  Fortunately, we obtain the above
value using renormalization group calculation~\cite{MKV:MHDRG}.  It is
interesting to note from Fig.~\ref{fig:et_fluid} that the temporal
evolution of energy does not clearly tell us which $\nu^*$ is the most
appropriate for LES.  Hence we should be careful in concluding the
appropriateness of $\nu^*$ using energy evolution.  The energy
spectrum has more information, and can provide us clues on the correct
choice of $\nu^*$.

From Fig.~\ref{fig:ek_fluid} we obtain the numerical value of $K$ to be
$1.7 \pm 0.1$; this value is close to the theoretically
calculated value 1.6 \cite{McCoWatt,MKV:MHD_PRE}.  Hence, the renormalized
viscosity predicted by Verma~\cite{MKV:MHD_PRE} appears to be
consistent and provides us a very good scheme for LES.

For LES of magnetohydrodynamic (MHD) turbulence, Agullo et
al.~\cite{MullCara_pop1}, and M\"{u}ller and Carati
\cite{MullCara_pop2,MullCara_cpc} applied dynamic gradient-diffusion
subgrid model.  The forms of eddy-viscosity and eddy-resistivity are derived
using dimensional arguments, but the constants are calculated using
dynamical LES procedure.  Their results match very well with DNS
counterpart.  In one of their main models, turbulent viscosity $\nu_t
\approx \bar{l}^{4/3} (\epsilon^K)^{1/3}$ and turbulent resistivity
$\eta_t \approx \bar{l}^{4/3} (\epsilon^M)^{1/3}$, where $\bar{l}$ is
the resolvable length scale on the LES grid, and $\epsilon^K$ and
$\epsilon^M$ are kinetic and magnetic energy dissipation applied by
the subgrid scale respectively.  Zhou et al.~\cite{Zhou:MHD_LES} have
studied subgrid scale and backscatter model for MHD turbulence using
EDQNM closure scheme.  Verma \cite{MKV:MHD_PRE,MKV:MHDRG} has also
calculated the above parameters using renormalization group procedure.
Simple calculations show that turbulent dissipative parameters of
Verma differ significantly from those of Agullo et
al.~\cite{MullCara_pop1} and M\"{u}ller and Carati
\cite{MullCara_pop2,MullCara_cpc}, as well as from those of Zhou et
al.~\cite{Zhou:MHD_LES}.  In the following discussions we will compare
the LES results from the above three approaches.

For MHD turbulence we apply the same LES method as described for fluid
turbulence using renormalized parameters.  The pseudo-spectral method
to solve MHD equations is very similar to that of fluid turbulence.
We also confine ourselves to zero cross helicity, i.e., (${\bf u \cdot
b}=0$), and zero mean magnetic field.  The difference of LES and DNS
is in the values of viscosity and resistivity.  In DNS we take
$\nu_0=0.00015$ and $\eta_0=0.00015$ with hyperviscosity and
hyperesistivity parameters $k_{eq}=7$. However, in LES we take
$\nu(k_C)=\nu_{r}(k_C)$, and $\eta(k_C)=\eta_{r}(k_C)$, where $k_C$ is
the cutoff wavelength.  The renormalized viscosity $\nu_r(k_C)$, and
renormalized resistivity $\eta(k_C)$ are taken from Verma
\cite{MKV:MHD_PRE,MKV:MHDRG} as
\begin{eqnarray}
\nu_r(k_C) & = & (K^u)^{1/2} \Pi^{1/3} k_C^{-4/3} \nu^*    
         \label{eqn:nuk} \\
\eta_r(k_C) & = & (K^u)^{1/2} \Pi^{1/3} k_C^{-4/3} \eta^*  .
        \label{eqn:etak}
\end{eqnarray}
Here $K^u$ is Kolmogorov's constant for MHD, $\Pi$ is the total energy
flux, and $\nu^*, \eta^*$ are renormalized parameters.  The parameters
$\nu^*, \eta^*$, and $K^u$ depend on the Alfv\'{e}n ratio $r_A$, which
is the ratio of kinetic and magnetic energy.  In our decaying MHD
turbulence simulation, we start with unit total energy and $r_A=8.0$.
The ratio of magnetic to kinetic energy grows as a function of time as
expected.  Therefore, we need to compute the renormalized parameters
for various values of $r_A$.  The parameters have been calculated
using the procedure described in Verma~\cite{MKV:MHDRG}, and they are
shown in Table 1.  We use the appropriate $\nu^*$ and $\eta^*$
given in the table for our simulations.  The energy cascade rates are
computed using Fast Fourier Transforms~\cite{fftw}.  We take
$\nu_{r}(k_C)$ and $\eta_{r}(k_C)$ from Eqs.~(\ref{eqn:nuk},
\ref{eqn:etak}).  The energy flux $\Pi$ changes with time; we compute
$\Pi$ dynamically every $0.01$ time-unit.
We carried out LES for MHD up to 25 nondimensional time units, and it
took approximately 55 hours.

\begin{figure}[h]
\centerline{\mbox{\psfig{file=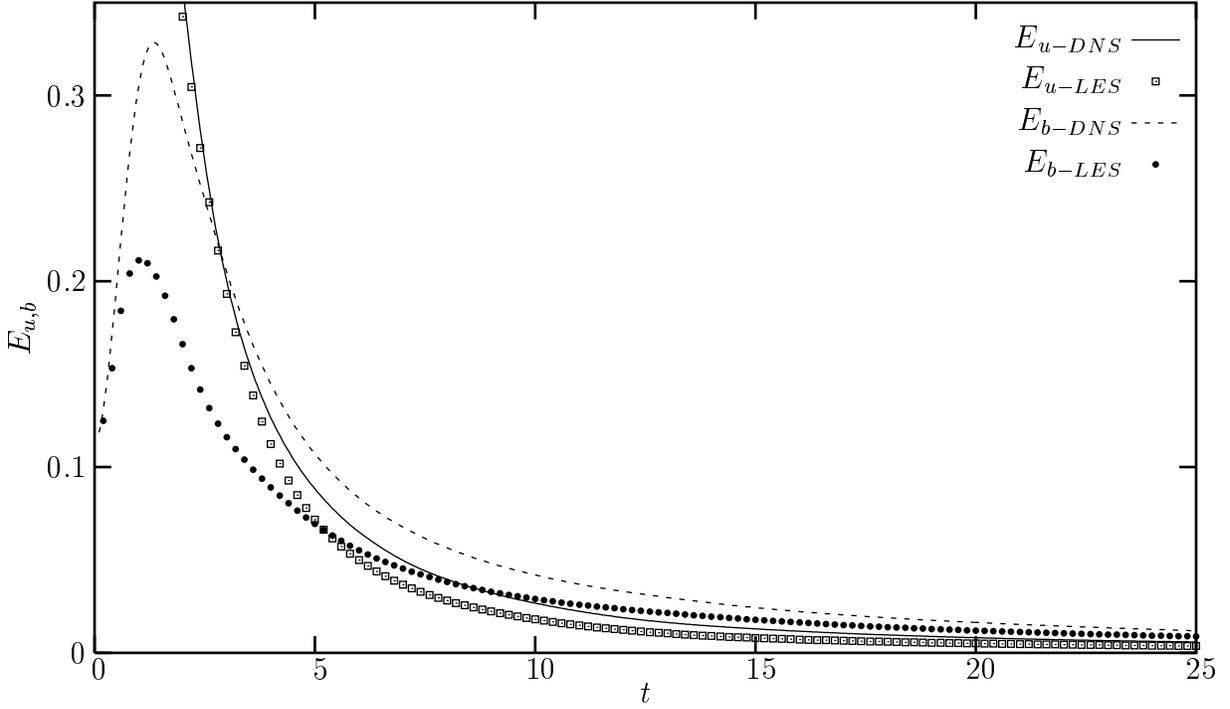,width=1.0\textwidth}}}
\caption{Temporal evolution of total kinetic and magnetic energy in
MHD turbulence using DNS and LES.  The kinetic energy matches quite
well in both the schemes, but magnetic energy evolves somewhat
differently.}
\label{fig:et_mhd}
\end{figure}

\begin{figure}[h]
\centerline{\mbox{\psfig{file=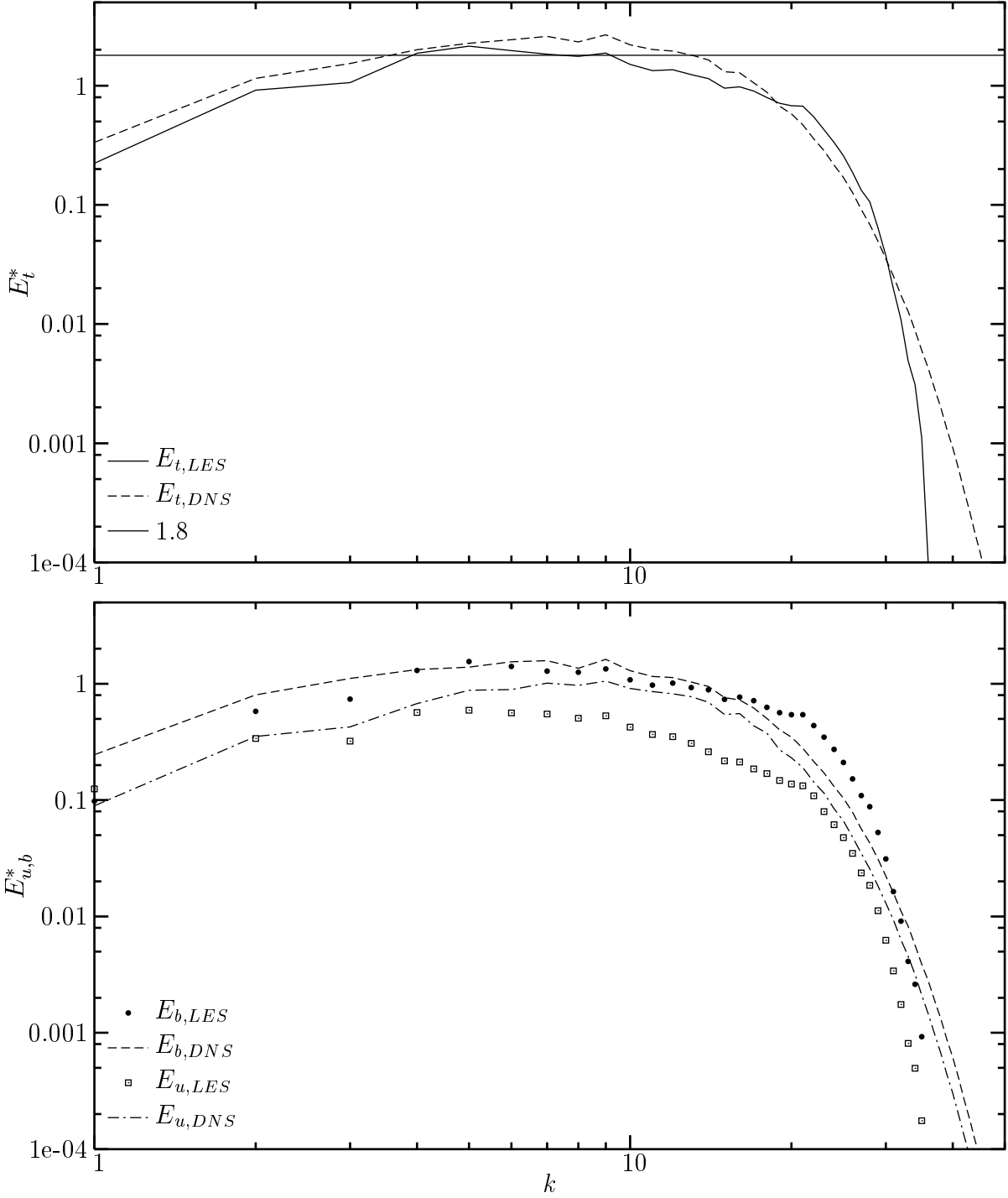,width=1.0\textwidth}}}
\caption{Plots of normalized spectra
$E'(k)=E(k)k^{5/3}\epsilon^{-2/3}$ with wavenumber $k$ for MHD
turbulence. The straight line shows the value of $K_o$ for LES run.}
\label{fig:ek_mhd}
\end{figure}

The evolution of kinetic and magnetic energies are shown in
Fig.~\ref{fig:et_mhd} as a function of time.  The evolution of kinetic
energy using LES is quite close to that using DNS.  However, the
evolution of magnetic energy does not match very well.  Comparatively,
LES of Agullo et al.~\cite{MullCara_pop1} and M\"{u}ller and Carati
\cite{MullCara_pop2,MullCara_cpc} yield a better fit to the temporal
evolution of energy.  Fig.~\ref{fig:ek_mhd} shows the energy
spectra of kinetic and magnetic energies for $r_A=0.5$ at 27 time
units of DNS and 12 time units of LES. We find that the energy spectra
calculated in LES matches quite well with that in DNS.  The
Kolmogorov's constant as indicated by the straight line in upper part
of Fig.~\ref{fig:ek_mhd} is found to be $1.8 \pm 0.2$, which is
close to the theoretical value calculated in
\cite{MKV:MHD_PRE,MKV:MHDRG}.  We conclude that the LES based on
renormalized parameters of Verma \cite{MKV:MHD_PRE,MKV:MHDRG} is quite
good.  Our numerical results are comparable with results of Agullo et
al.~\cite{MullCara_pop1} and M\"{u}ller and
Carati~\cite{MullCara_pop2,MullCara_cpc}.  However, we believe that
our parameters, which are based on field-theoretic calculations, are
on a somewhat stronger footing as compared to those used in earlier
LES methods.

To conclude, we have devised a LES procedure for fluid and MHD
turbulence in Fourier space using the renormalized parameters.  We
take renormalized parameters from Verma~\cite{MKV:MHD_PRE,MKV:MHDRG}
and carry out LES for $64^3$ grid.  When LES results are compared with
DNS of size $128^3$ with the same initial conditions, we find that our
LES results on energy evolution and spectra match quite well with the
DNS results, except for the temporal evolution of magnetic energy.
The inertial range of LES is much larger compared to DNS of the same
size.  Our results shows that substitution of renormalized parameters
for eddy viscosity in LES yield excellent results.  Hence, we
demonstrate the usefulness of renormalized parameters in LES
calculations.


\pagebreak

\centerline{\large Figure Captions}
\vspace{1cm}

\noindent
{\bf Fig. 1} \, Temporal evolution of energy in fluid turbulence using DNS
and LES.  The figure contains Energy($E$) vs time plots for DNS128
(solid line), DNS (DNS64), and three LES runs using $\nu^*$ equal to
0.38 (LES1), 0,25 (LES2) and 0.48 (LES3).  The evolution in LES for
all the three $\nu^*$ is quite close to DNS128, but not to
DNS64. \\

\noindent
{\bf Fig. 2} \, Energy spectrum for fluid turbulence is calculated using DNS
and LES.  The figure contains plots of normalized energy spectrum
$E'(k)=E(k)k^{5/3}\epsilon^{-2/3}$ with wavenumber $k$ after 50 time
units for DNS128, DNS64, and three LES runs using $\nu^*$ equal to 0.38 (LES1),
0,25 (LES2) and 0.48 (LES3).  We get the best inertial range for
$\nu^*=0.38$.  The Kolmogorov's constant is found to be $1.7 \pm
0.1$. DNS64 run has hardly any inertial range. \\

\noindent
{\bf Fig. 3} \, Temporal evolution of total kinetic and magnetic energy in
MHD turbulence using DNS and LES.  The kinetic energy matches quite
well in both the schemes, but magnetic energy evolves somewhat
differently. \\

\noindent
{\bf Fig. 4} \, Plots of normalized spectra
$E'(k)=E(k)k^{5/3}\epsilon^{-2/3}$ with wavenumber $k$ for MHD
turbulence. The straight line shows the value of $K_o$ for LES run. \\

\pagebreak

\begin{table}
\caption{\label{tab:table1}The values of renormalized parameters for 
viscosity ($\nu^*$) and resistivity  $\eta^*$ in MHD turbulence at 
various values of Alfv\'{e}n ratio $r_A$ and zero cross helicity.  We also
list the Kolmogorov's constant $K^u$ for MHD turbulence.}

\begin{ruledtabular}
\begin{tabular}{cccc}
$r_A$ & $\nu^*$ & $\eta^*$ & $K^u$ \\
\hline
0.3 & 7.20 & 0.20 & 0.50 \\
0.4 & 3.15 & 0.38 & 0.53 \\
0.5 & 2.08 & 0.50 & 0.55 \\
0.6 & 1.64 & 0.57 & 0.59 \\
0.7 & 1.38 & 0.61 & 0.63 \\
0.8 & 1.21 & 0.64 & 0.67 \\
0.9 & 1.09 & 0.67 & 0.71 \\
1.0 & 1.00 & 0.69 & 0.75 \\
2.0 & 0.65 & 0.77 & 1.01 \\
3.0 & 0.54 & 0.79 & 1.15 \\
4.0 & 0.49 & 0.81 & 1.23 \\
5.0 & 0.47 & 0.82 & 1.28 \\
\end{tabular}
\end{ruledtabular}
\end{table}

\pagebreak

%
%

\newpage

\newpage

\newpage

\newpage

\end{document}